\documentclass[aps,prb,twocolumn,superscriptaddress]{revtex4-1}

\usepackage{bm}        
\usepackage{amsmath,amssymb,amsthm}   
\usepackage{mathtools}
\usepackage{tabularx}
\usepackage{array}
\usepackage{color,soul}
\usepackage{diagbox}

\usepackage{hyperref}
\usepackage{grffile}
\usepackage{graphicx}
\usepackage{float}
\usepackage{chemformula}
\usepackage{gensymb}
\usepackage{multirow}

\graphicspath{{./figures/}}

\newcommand{\wn}{\,cm$^{-1}$}
\newcommand{\Ag}[1]{A$_\mathrm{g}^{#1}$}
\newcommand{\Eg}[1]{E$_\mathrm{g}^{#1}$}
\newcommand{\tmag}{{two-magnon}}
\newcommand{\TN}{{$T_\textrm{N\'eel}$}}

\begin{document}
\title{Magnon-phonon hybridization in quasi-2D antiferromagnet \ch{MnPSe_3}}
\author{Thuc T. Mai}
\email{thuc.mai@nist.gov}
\affiliation{Nanoscale Device Characterization Division, Physical Measurement Laboratory, NIST, Gaithersburg, MD}
\author{K.F. Garrity}
\email{kevin.garrity@nist.gov}
\affiliation{Materials Measurement Science Division, Materials Measurement Laboratory, NIST, Gaithersburg, MD}
\author{A. McCreary}
\affiliation{Nanoscale Device Characterization Division, Physical Measurement Laboratory, NIST, Gaithersburg, MD}
\author{J. Argo}
\affiliation{Department of Materials Science and Engineering, Ohio State University, Columbus, OH}
\author{J.R. Simpson}
\affiliation{Nanoscale Device Characterization Division, Physical Measurement Laboratory, NIST, Gaithersburg, MD}
\affiliation{Physics, Astronomy, and Geosciences, Towson University, Towson, MD}
\author{V. Doan-Nguyen}
\affiliation{Department of Materials Science and Engineering, Ohio State University, Columbus, OH}
\author{R. Vald\'es Aguilar}
\affiliation{Center for Emergent Materials, Department of Physics, The Ohio State University. Columbus, OH}
\author{A.R. Hight Walker}
\email{angela.hightwalker@nist.gov}
\affiliation{Nanoscale Device Characterization Division, Physical Measurement Laboratory, NIST, Gaithersburg, MD}
\date{\today}
\begin{abstract}
Magnetic excitations in van der Waals (vdW) materials, especially in the two-dimensional (2D) limit, are an exciting research topic from both the fundamental and applied perspectives. Using temperature-dependent, magneto-Raman spectroscopy, we identify the hybridization of two-magnon excitations with two separate E$_\mathrm{g}$ phonons in \ch{MnPSe_3}, a magnetic vdW material that could potentially host 2D antiferromagnetism. Results from first principles calculations of the phonon and magnon spectra further support our identification. The Raman spectra's rich temperature dependence through the magnetic transition displays an avoided-crossing behavior in the phonons' frequency and a concurrent decrease in their lifetimes. We construct a model based on the interaction between a discrete level and a continuum that reproduces these observations. The strong magnon-phonon hybridization reported here highlights the need to understand its effects on spin transport experiments in magnetic vdW materials.
\end{abstract}
\maketitle{}



Magnons, the quantized magnetic excitations in solids, have been the subject of a plethora of exciting research in recent years. From the quantum computing perspective, magnons have been shown to couple to a superconducting qubit in a cavity\cite{Tabuchi_2015,Lachance-Quirione_2017} to make up the field of quantum magnonics \cite{Tabuchi_2016,Chumak_2017,Lachance_Quirion_2019,Clerk2020}. Magnons have also received attention from the spintronics community due to their ability to transport spin angular momentum efficiently\cite{Stamps_2014,Chumak_2017,Bozhko_2020}. Recently discovered magnetic 2D materials can retain intrinsic magnetism down to the monolayer limit\cite{McGuire_2017,Susner_2017,Burch_2018_Review,Gibertini2019,Cortie2DReview2020}. Magnons have now been observed in several materials of this class, including ferromagnetic (FM) magnons in \ch{CrI_3}\cite{Lee2020,Cenker2020,Li2020,McCreary2020CrI3} and antiferromagnetic (AFM) magnons in both $\alpha$-\ch{RuCl_3}\cite{Banerjee_2017,Little_2017,Wang_2017,Wu_2018,Wulferding2020,Ponomaryov_2020,Sahasrabudhe_2020} and the \textit{M}P\textit{X}$_3$ family (where \textit{M} is Fe, Mn, or Ni and \textit{X} is S or Se)\cite{Wildes_1998,Kim2019_NiPS3,McCreary_2020_PRB,calder2020magnetic}. These observations have resulted in an increased interest in using magnetic 2D materials for magnon spintronics. 

Along these lines, several experimental studies have explored the capability of \ch{CrI_3} and \ch{Fe_3GeTe_2} as a tunable spin filter\cite{Klein2018,Jiang2018,Wang2019,Albarakati2019}. Additional experiment on \ch{MnPS_3} has provided evidence of magnon-mediated spin transport\cite{Xing2020}. Furthermore, magnon transport over macroscopic scales has been linked to the hybridization between magnons and acoustic phonons\cite{An_2020,Ruckriegel_2020_PRL}. The same hybridization could theoretically lead to topologically non-trivial quasiparticle bands\cite{Takahashi2016,Zhang_2020_SU3}. However, the interactions between magnons and phonons and their hybridization are generally unfavorable as they can cause an increase in scattering rate and a decrease in quasiparticle lifetime. Understanding magnon-phonon interactions is essential for applications that rely on preserving the magnon coherence over extended length and time scales.

In this letter, we use a novel combination of temperature and magnetic field-dependent Raman spectroscopy and first principles calculations to investigate the magnon-phonon interactions in the AFM vdW material \ch{MnPSe_3}. We identify the hybridization between a continuum of two-magnon excitations and two separate Raman-active phonons as a function of temperature. A model with magnon-phonon interaction is constructed to reproduce our experimental results. We find that a broad range of \tmag\ excitations are hybridized with the two Raman-active phonons over a wide range of temperatures. This interaction causes a dramatic reduction in quasiparticle lifetimes in the system, impacting potential magnon spintronics devices.

Bulk crystals of \ch{MnPSe_3} are synthesized via the vapor transport method\cite{SI}. The sample is mounted inside a closed-cycled magneto-optical cryostat and immersed in He exchange gas. A static magnetic field can be applied either normal or parallel to the vdW layers. Raman spectra are measured using a triple grating spectrometer and in the backscattering geometry, where a 515\,nm Ar$^+$ laser was used to excite an area of approximately 1\,$\mu m^2$ on the sample. A combination of linear polarizers and half-wave plates are used to select parallel (VV) and crossed (VH) polarization configurations with respect to the incoming and scattered light, as well as to correct for Faraday rotation in the objective when the static field is normal to the plane of the sample. Due to the relatively small momentum that the laser photons can impart on the crystal, the excitations in a Raman scattering experiment are generally limited to $q \approx 0$.

\begin{figure}[th]
\includegraphics[width=.9\columnwidth]{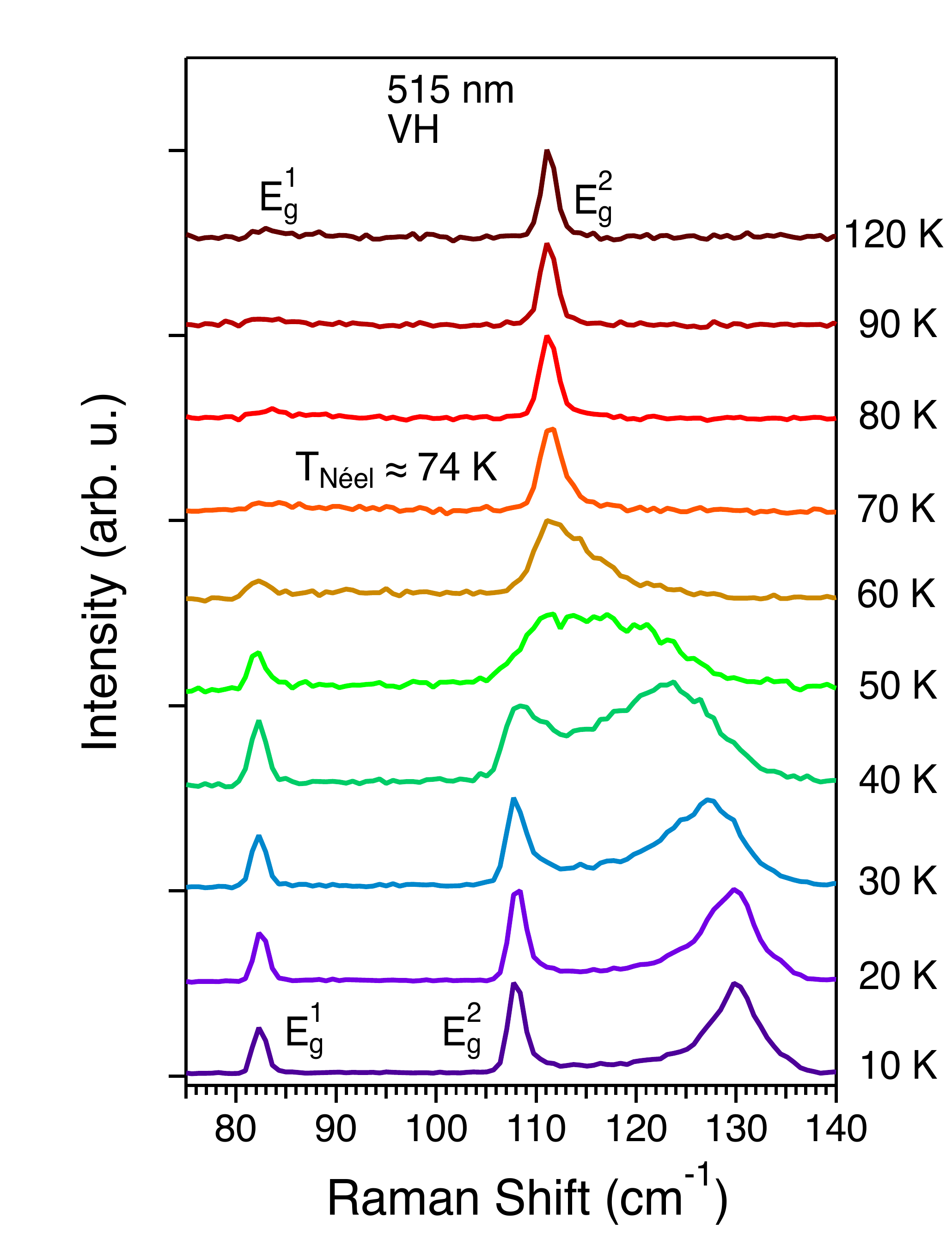}
\caption{\textbf{Temperature dependent Raman spectra of \ch{MnPSe_3}}. The Raman spectra for crossed (VH) polarization collected through the N\'eel transition at 74\,K. The spectra are normalized to the peak intensity of \Eg{2}, and offset for clarity. The two Raman phonon modes are labeled with the irreducible representation of the point group $\overline{3}$. Below \TN, a new scattering peak appears between 115\wn\ - 130\wn, while the \Eg{1} and \Eg{2} phonons undergo changes in both intensity and frequency.
}
\label{fig:Phonons}
\end{figure}

First principles density functional theory (DFT)\cite{hk,ks} calculations are carried out using the Quantum Espresso code\cite{qe}, with GBRV pseudopotentials\cite{gbrv}. We utilize the PBEsol exchange correlation function functional\cite{pbesol} and a Hubbard U correction of 4 eV on the Mn-d states (DFT+U)\cite{DFTU}. A discussion of the effects of different U values is included in the Supplementary Information (SI)\cite{SI}. The phonon and magnon energies are calculated using finite differences approaches\cite{spring_cluster, phonopy}.

The \textit{M}\ch{PSe_3} family of materials shows higher magnetic anisotropy in the bulk \cite{Bhutani_2020,Jeevanandam_1999} than its sulfide counterpart, particularly when Mn or Fe occupies the transition metal site \textit{M}. Such anisotropy is a necessary ingredient to stabilize the long-range magnetic ordering in the 2D limit\cite{Burch_2018_Review,Gibertini2019,Cortie2DReview2020}. The crystal symmetry of \ch{MnPSe_3} is rhombohedral at room temperature, with space group R$\overline{3}$, and remains so below \TN. Previous elastic neutron scattering experiments have determined that \ch{MnPSe_3} is a N\'eel type antiferromagnet within each vdW plane below 74\,K\cite{WIEDENMANN_1981,calder2020magnetic}. The spin $\frac{5}{2}$ Mn ions are arranged in a honeycomb lattice, with the magnetic moments pointing nearly parallel to the honeycomb plane\cite{WIEDENMANN_1981,calder2020magnetic}. 

We measure the polarized Raman spectra of \ch{MnPSe_3} in its paramagnetic phase, and characterize the symmetries of the Raman phonon modes with the assistance from DFT\cite{SI}. As the temperature is lowered through \TN\ (Fig. \ref{fig:Phonons}), several striking features appear in the frequency region between 75\,\wn\ and 140\,\wn: (1) the \Eg{1} and \Eg{2} phonons undergo significant changes in intensity and frequency and (2) a new scattering intensity peak appears to split off from \Eg{2} and continues to increase in frequency to approximately 130\wn\ at 10\,K.

\begin{figure}[bh!]
\includegraphics[width=1.03\columnwidth]{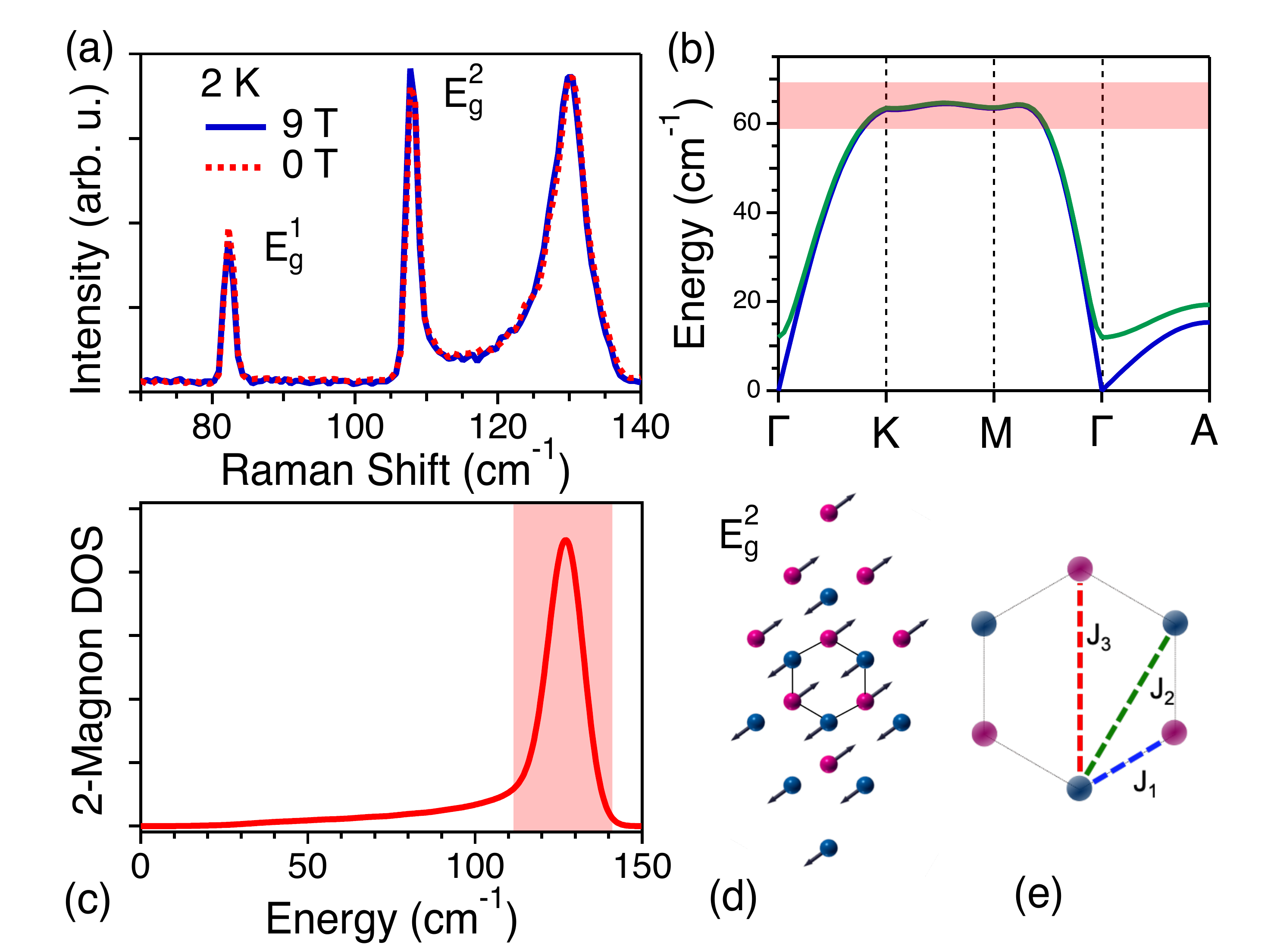}
\caption{\textbf{Two-magnon scattering} (a) The scattering intensity around 130\wn\ shows no discernible difference between 0\,T and 9\,T applied either parallel and perpendicular to the honeycomb plane\cite{SI}. (b) The magnon dispersion with easy-plane anisotropy resulting from linear spin wave theory with highlighted zone-edge magnon states. The gapless (blue) and gapped (green) bands correspond to in-plane and out-of-plane spin excitations, respectively. (c) The calculated two-magnon density of states, $DOS_{2M}$, with its highlighted peak at twice the one-magnon energies at the Brillouin zone boundary (also highlighted in (b)), corresponds to pairs of magnons with equal and opposite momentum. (d) Top view of the normal mode vibrations for \Eg{2}. The Mn atoms are shown here with different colors to represent different spin directions. The arrows represent the directions of the atomic vibrations. (e) The Mn$^{2+}$ exchange coupling between nearest-neighbor ($J_1$), next-nearest-neighbor ($J_2$), third-nearest-neighbor ($J_3$).
}
\label{fig:Theory}
\end{figure}

The new mode, with \Eg{} symmetry\cite{SI}, cannot be attributed to a phonon according to the DFT calculations. This mode is asymmetric, with a cutoff at high frequency, a peak intensity around 130\wn, and a low-intensity tail that merges with the \Eg{2} phonon (Fig. \ref{fig:Theory}a). A previous study interpreted the origin of the 130\wn\ peak as one-magnon scattering\cite{Makimura_1993} based on its strong frequency shift as a function of temperature. A one-magnon scattering mode in an AFM is expected to split in an applied magnetic field due to its net magnetic moment, $\Delta S = \pm 1$\cite{FleuryLoudon}. Such behavior has been observed in three-dimensional AFMs such as \ch{MnF_2} and \ch{FeF_2}\cite{FleuryLoudon,Rezende_AFMR}, and more recently in vdW AFMs\cite{McCreary_2020_PRB,McCreary2020CrI3,Cenker2020,Li2020}. However, when we measure the low temperature ($T\,=2\,$K) Raman spectra as a function of static applied magnetic field, both parallel and perpendicular to the honeycomb plane, we find no change in the 130\wn\ peak (Fig. \ref{fig:Theory}(a) and SI\cite{SI}). This result strongly suggests that the peak at 130 \wn\ is not due to one-magnon scattering. Instead, we propose that its origin is due to \tmag\ scattering.

Qualitatively, pairs of magnon states with opposite momenta and opposite spins are excited by the Raman scattering process, conserving both linear and angular momentum\cite{FleuryLoudon}. The picture of opposite spin magnons, with $\Delta S$ = 0, would explain the lack of dependency of the two-magnon scattering on the applied magnetic field. We use a Heisenberg Hamiltonian model to further support our interpretation of the 130 \wn\ peak as two-magnon scattering. The Hamiltonian is comprised of spin exchange coupling terms, $J_{ij}$, and on-site anisotropy, $D$:
\begin{equation} \label{Hamiltonian}
\mathcal{H} = \frac{1}{2}\sum_{i\neq j} J_{ij} \mathbf{S}_i \cdot \mathbf{S}_j  \;+\;   D \sum {S^z_i}^2.
\end{equation}

We obtain the parameters ($J_{ij}$, $D$) by performing a least squares fit to DFT energies with various magnetic configurations\cite{spring_cluster}, with the nearest-neighbor (NN) $J_1=0.758$\,meV, next-nearest-neighbor (NNN) $J_2=0.069$\,meV, and third-nearest-neighbor (TNN) $J_3=0.474$\,meV (see Fig. \ref{fig:Theory}e). The exchange coupling to adjacent the vdW planes is 0.002\,meV, 0.033\,meV, and 0.010\,meV for the 1$^\textrm{st}$, 2$^\textrm{nd}$, and 3$^\textrm{rd}$ nearest neighbor, respectively. The on-site anisotropy, $D$, is 0.046\,meV. The magnon bands are calculated using linear spin wave theory (LSWT). Fig. \ref{fig:Theory}b shows the magnon dispersion along the high symmetry points in the reciprocal lattice space: $\mathbf{\Gamma}=(0,0,0)$, \textbf{K} $=(1/3,1/3,0)$, \textbf{M} $=(1/2,0,0)$, and \textbf{A} $=(0,0,1/2)$.

As expected from the significantly weaker exchange interaction perpendicular to the vdW plane, the magnon dispersion along \textbf{A} is smaller than along the in-plane directions (Fig. \ref{fig:Theory}b). The easy-plane anisotropy creates a gapless magnon band (blue curve in Fig. \ref{fig:Theory}b) for in-plane $\Delta S$, and a gapped magnon band (green curve) for out-of-plane $\Delta S$. We did not observe the gapped magnon mode at $\mathbf{\Gamma}$ in our experiment. In a recent inelastic neutron scattering experiment on \ch{MnPSe_3}\cite{calder2020magnetic}, the measured magnon dispersion agrees with our LSWT results quantitatively, on both the zone-center and zone-edge magnon energies. The experimental $J_{ij}$ values are different from ours (for a more detailed discussion, see the SI\cite{SI}).

The \tmag\ density of states, $DOS_{2M}(\omega) = \sum_{i,{\bf k} } \delta(\omega - 2 \omega_{i,{\bf k}})$ with artificial linewidth broadening is shown in Fig. \ref{fig:Theory}c and qualitatively agrees with our observation of the broad 130 \wn\ peak (compare Fig. \ref{fig:Theory}a,c). The numerical value of the first principles \tmag\ peak depends on the choice of Hubbard U as we discuss in the SI\cite{SI}. Rather than a single sharp mode, the \tmag\ scattering intensity is due to a continuum of \tmag\ excitations. In our calculation, the $DOS_{2M}$ has its highest intensity around 130\wn. This energy corresponds to the single magnon states near the Brillouin zone edge at approximately 65\wn(near reciprocal points \textbf{K}, \textbf{M}), as highlighted in Fig. \ref{fig:Theory}b, c. The intensity cutoff at high frequency comes from the lack of magnon states above the zone-edge energy, while the weaker but broad scattering intensity tail receives contributions from pairs of lower energy magnons with opposite momenta. 

\begin{figure}[bh!]
\includegraphics[width=1.05\columnwidth]{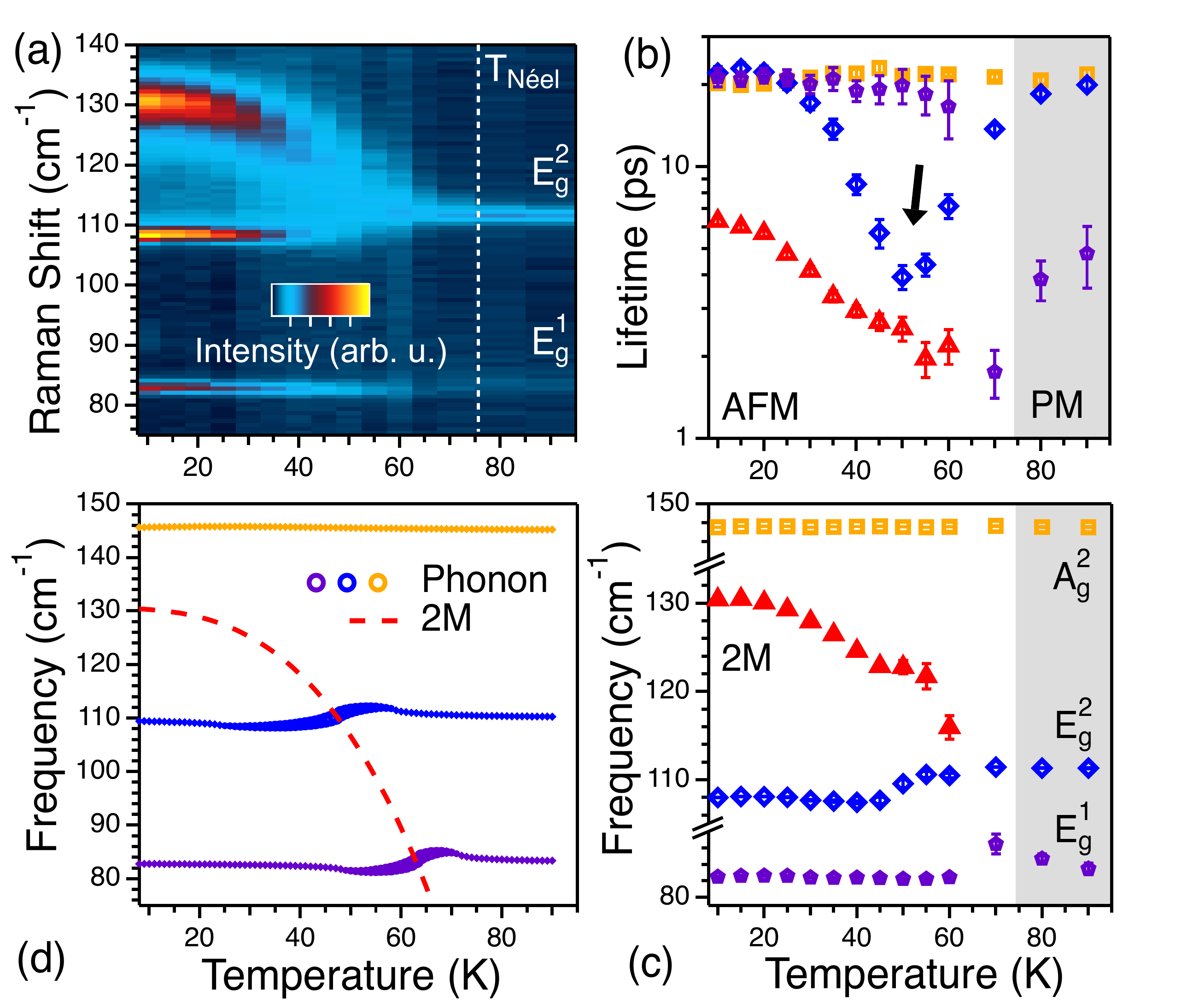}
\caption{\textbf{Phonon and Two-Magnon Hybridization}. (a) False color plot showing Raman intensity between 75\wn\ and 140\wn\ as a function of temperature through \TN. The phonons \Eg{1} and \Eg{2} suddenly decrease in frequency as temperature is lowered. The \tmag\ scattering appears below \TN\ and increases gradually in frequency to 130\wn at 10\,K. (b), (c) The results of fitting \Eg{1}, \Eg{2}, and the \tmag\ peak intensity (labeled 2M), in addition to \Ag{2}. The jumps in peak frequency of each phonon correlate to abrupt decreases in the quasiparticle lifetime (inversely proportional to the scattering linewidth), which occur around \TN\ for \Eg{1} and 55\,K for \Eg{2} (highlighted by the black arrow). \Ag{2} shows no discernible change in this temperature range. The error bars represent one standard deviation from the best fit values. (d) Frequency versus temperature results using a Fano-like model that take into account a broad scattering feature, with its center labeled by 2M, and its interaction with discrete states (phonons) as a function of temperature. The size of circles represents the line width.
}
\label{fig:Tdep}
\end{figure}

A detailed look at the frequency range between 75\wn\ and 140\wn\ as the temperature is varied from 100\,K to 10\,K is shown in Fig. \ref{fig:Tdep}, including the \Eg{1}, \Eg{2}, and \Ag{2} phonons and the \tmag\ excitations. The false-color plot (Fig. \ref{fig:Tdep}a) shows dramatic changes in the frequency and intensity of \Eg{1} and \Eg{2} as a function of temperature, along with the appearance of the \tmag\ excitations below 60\,K, which increases in frequency to approximately 130\wn\ at 10\,K. The \tmag\ excitations are of $\rm{E_g}$ symmetry, appearing in both parallel and crossed polarization configurations\cite{SI}. Further supporting our assignment of the feature at 130\wn\ is its similarity to the \tmag\ excitations measured in \ch{FeF_2} and \ch{MnF_2}. In both materials, the \tmag\ excitations shows a strong decrease in frequency as temperature increases to \TN\cite{Cottam_1983,LockwoodCottom1987}. 

Fitting results for the peak positions and lifetimes for \Eg{1}, \Eg{2}, \Ag{2}, and the \tmag\ (2M) are shown in Fig. \ref{fig:Tdep}b, c (see SI for more fitting details\cite{SI}). The observed shifts in frequency for \Eg{1} and \Eg{2} below \TN\ are larger than any other phonons in the spectra, as evidenced in the comparison with \Ag{2} (Fig. \ref{fig:Tdep}c). Immediately below \TN, we see an abrupt decrease in the frequency of \Eg{1} by more than 1\wn (Fig. \ref{fig:Tdep}c). Concurrent with the change in frequency, the lifetime of \Eg{1} decreases sharply (Fig. \ref{fig:Tdep}b). Such a dramatic jump is indicative that \Eg{1} has coupled with another state or states. Although \Eg{2} exhibits the same behavior as \Eg{1}, the onset occurs at a noticeably lower temperature of approximately 55\,K. These behaviors are reminiscent of an avoided-crossing phenomenon, with a few key differences, as detailed below.

While the lifetimes of the \Eg{1} and \Eg{2} phonons recover to approximately the same value as other phonons at 10\,K, the peak frequencies remain below their high temperature values (Fig. \ref{fig:Tdep}b). We propose that the temperature dependence of \Eg{1} and \Eg{2} is caused by their hybridization with the strongly temperature dependent \tmag\ excitations. At \TN\ $\approx$ 74\,K, the \tmag\ excitations are at a lower frequency than both phonon modes, and thus there is no overlap. As the \tmag\ peak increases in frequency with decreasing temperature, it first crosses path with \Eg{1} slightly below \TN, and then crosses with \Eg{2} near 55\,K. When the phonon modes overlap with the \tmag\ excitations, they become broad and asymmetric, and their center frequencies shift away from the \tmag\ peak intensity.

Since there is an overlap in energy, the \tmag\ excitations and the \Eg{1}, \Eg{2} phonons can interact with each other \textit{via} the exchange striction effect\cite{Baltensperger1968,Cheong_2007}. The normal mode motion of \Eg{2} phonon shows the modulation in distance between NN and TNN Mn$^{2+}$ atoms (Fig. \ref{fig:Theory}d). These modulations would affect the superexchange coupling due to the change in various bond lengths and angles. This coupling results in strong hybridization between the phonons and the \tmag\ excitations when their energies are degenerate.

\begin{figure}[th!]
\includegraphics[width=\columnwidth]{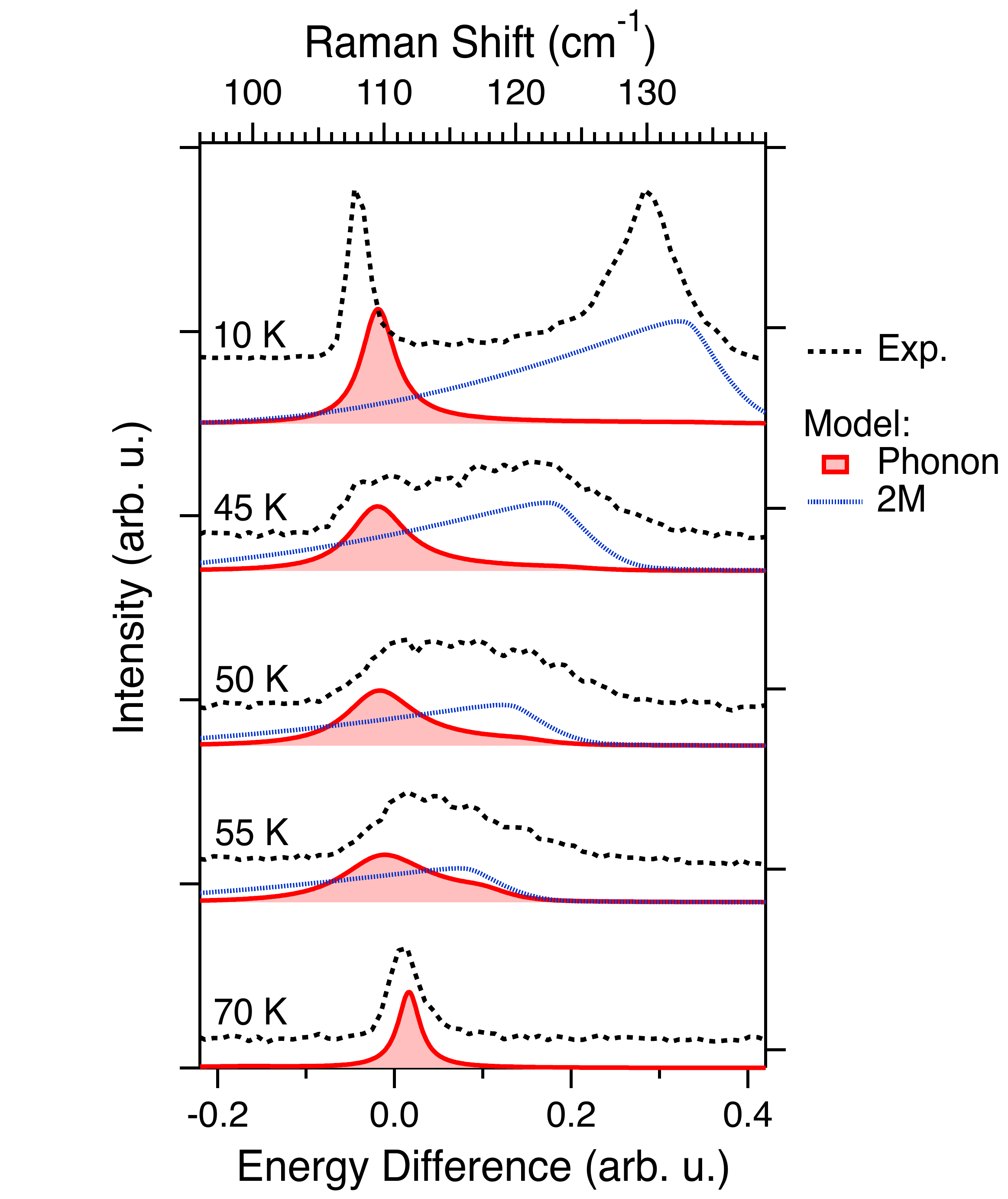}
\caption{\textbf{Hybridization model}. Calculated scattering intensity due to the two-magnon (2M, blue dash) continuum and a discrete phonon state (solid red filled). The relative energy between the 2M and phonon simulates different temperatures. Select experimental spectra (black dash) from crossed polarization (VH) are chosen to compare to the model. Significant hybridization can be seen in the 55\,K and 50\,K spectra by the deviation of the lineshape from a simple Lorentzian.
}
\label{fig:hybrid}
\end{figure}

To analyze this interaction more quantitatively, we construct a model based on the work by Fano\cite{Fano1961} on the interaction of a discrete resonance (phonon) with a continuum (\tmag). The model consists of a broadly peaked, temperature-dependent \tmag\ continuum that interacts with the narrow, temperature-independent $\mathbf{\Gamma}$-point phonons. We assume the squared of the interaction strength is proportional to the intensity of the continuum peak (see SI for more details\cite{SI}). By assuming a simple, functional form for the continuum, we solve the model analytically for weak coupling or numerically for stronger coupling. The main results are presented in Fig. \ref{fig:Tdep}d, which shows the temperature dependent phonon frequencies, with the linewidths shown as circle sizes. 

We find that this simple model reproduces several otherwise puzzling features of the experiment, shown in Fig. \ref{fig:Tdep}a-c. (1) The frequencies of the \Eg{1} and \Eg{2} modes are repelled by the peak intensity in the \tmag\ excitations, with the shift extending beyond the temperature region where the \tmag\ directly overlaps with the phonons. The direction of this frequency shift abruptly reverses sign when the peak of the \tmag\ excitations passes through the phonon, but only slowly returns to zero (as seen in \Eg{1} in Fig. \ref{fig:Tdep}c,d). (2) The phonon lifetimes drop significantly when the phonons overlap with \tmag\ excitations, but recover to their previous values outside of the overlap region. (3) The phonon modes become asymmetric, with a Fano-lineshape, when they overlap with the \tmag\ excitations. (4) Finally, phonon modes that never overlap with the \tmag\ excitations, such as the \Ag{2} mode at 149\,\wn, are nearly temperature-independent.

The interaction with a \tmag\ continuum explains the observation that the frequencies of \Eg{1} and \Eg{2} remain lowered after the \tmag's apparent peak crosses through. In a typical avoided-crossing phenomenon, one would expect the interaction to lift the degeneracy in the vicinity of the crossing, resulting in the hybridization of the two levels and an avoided-crossing. However, away from the original crossing point, the character of these discrete levels would be expected to return to their unhybridized state. A comparison between experimental spectra at 10\,K, 45\,K, 50\,K, 55\,K, and 70\,K and our model results are shown in Fig. \ref{fig:hybrid}. The strongest hybridization is shown in the phonon excitation channel that corresponds to the 55\,K experimental spectrum. The hybridization effect becomes less prominent at lower temperatures but still presents as a shift in the phonon energy. From our model and observations, we infer that the phonons remain hybridized with the \tmag\ excitations over a wide range of temperatures below \TN. Since the \tmag\ continuum covers a broad range of frequencies, the bare phonon frequency is always degenerate with some of the \tmag\ excitations. Consequently, the phonons \Eg{1}, \Eg{2}, and a broad range of states in the \tmag\ continuum are hybridized, leading to a reduction in the lifetimes of the quasiparticles involved, phonons and magnons.

A similar magnon spectrum and exchange parameters have been measured in \ch{MnPS_3}\cite{Wildes_1998}, a material with analogous crystallographic and magnetic structure to \ch{MnPSe_3}. In the sulfur compound, however, the weaker spin-orbit coupling of the sulfur anions produces an on-site anisotropy that favors the Mn$^{2+}$ spins to point out of the vdW plane. Furthermore, an abrupt decrease in phonon frequency as a function of temperature has been measured in the Raman spectrum of \ch{MnPS_3} down to a single layer thickness\cite{Kim_2019,Sun2019}. The same hybridization phenomenon between a phonon and \tmag\ excitations, as discussed herein, is likely to be responsible for the frequency shift in \ch{MnPS_3} even though the two-magnon excitations are not seen. 

Interestingly, an experiment using \ch{MnPS_3} as the spin carrying channel has provided evidence of magnon-mediated spin transport\cite{Xing2020}. In the study, spin angular momentum was carried by magnons over a length scale of several microns, demonstrating the material capability in magnon spintronics applications. It is still unknown as to how the magnon-phonon hybridization studied herein would impact spintronic devices. Our results pave the way for further examinations of the role of magnon-phonon hybridization on magnon coherence in spin transport experiments within the \ch{MPX_3} family of materials.

In summary, using a novel temperature and magnetic field-dependent Raman spectroscopy in the AFM phase of \ch{MnPSe_3}, we accurately assign the origin of the mode around 130\wn as scattering from a \tmag\ continuum of excitations. Using LSWT, we calculate the magnon spectrum and the \tmag\ DOS to support our assignment. The magnon calculation agrees well with the latest published experimental results\cite{calder2020magnetic}. The temperature-dependent Raman scattering data reveal an avoided-crossing like behavior, suggesting that the Raman-active phonons \Eg{1,2} and the magnetic excitations have hybridized in the AFM phase. By measuring the phonons' lifetime, we identify this hybridization as a source of magnon lifetime reduction in the material. A Fano-like model is constructed, where we discovered that the shift in frequency of two Raman phonons over a wide range of temperatures is due to their hybridization with the \tmag\ continuum. This work highlights the identification and direct measurement of a source of magnon lifetime reduction in 2D magnetic materials.

\bibliography{MnPSe3References}
\end{document}


\title{Magnon-phonon hybridization in quasi-2D antiferromagnet \ch{MnPSe_3}: Supplementary Information}
\author{Thuc T. Mai}
\affiliation{Nanoscale Device Characterization Division, Physical Measurement Laboratory, NIST, Gaithersburg, MD}
\author{K.F. Garrity}
\affiliation{Materials Measurement Science Division, Materials Measurement Laboratory, NIST, Gaithersburg, MD}
\author{A. McCreary}
\affiliation{Nanoscale Device Characterization Division, Physical Measurement Laboratory, NIST, Gaithersburg, MD}
\author{J. Argo}
\affiliation{Department of Materials Science and Engineering, Ohio State University, Columbus, OH}
\author{J.R. Simpson}
\affiliation{Nanoscale Device Characterization Division, Physical Measurement Laboratory, NIST, Gaithersburg, MD}
\affiliation{Physics, Astronomy, and Geosciences, Towson University, Towson, MD}
\author{V. Doan-Nguyen}
\affiliation{Department of Materials Science and Engineering, Ohio State University, Columbus, OH}
\author{R. Vald\'es Aguilar}
\affiliation{Center for Emergent Materials, Department of Physics, The Ohio State University. Columbus, OH}
\author{A.R. Hight Walker}
\affiliation{Nanoscale Device Characterization Division, Physical Measurement Laboratory, NIST, Gaithersburg, MD}
\date{\today}
\maketitle{}
\section{Growth and Characterization}
We synthesized the samples via the vapor transport method. Finely ground powders of red phosphorus, manganese, and selenium were mixed in stoichiometric ratio for a total precursor weight of 1.52 g. To this mixture, 0.03 g of iodine was added to serve as a transport agent. The powders were transferred to a fused quartz tube 20 cm in length with inner diameter 13.5 mm and sealed under vacuum while cooling the tube with liquid N$_2$ to prevent loss of iodine. The sealed ampoule was heated in the center of a two-zone tube furnace, with the side containing the precursors at 650\,$^\circ$C and the cold end at 550\,$^\circ$C. The temperature was ramped up at 10\,$^\circ$C/min, held for one week, and reduced to 25\,$^\circ$C at a rate of 0.1\,$^\circ$C/min. The wine-colored crystal flakes were rinsed with acetone to remove excess iodine. X-ray diffraction of the crystals were collected using lab X-ray source ($\lambda$=1.540593 \AA) at 2\,$^\circ$C/minute scan rate in Bragg-Brentano geometry. 

\begin{figure*}[th]
\includegraphics[width=0.5\textwidth]{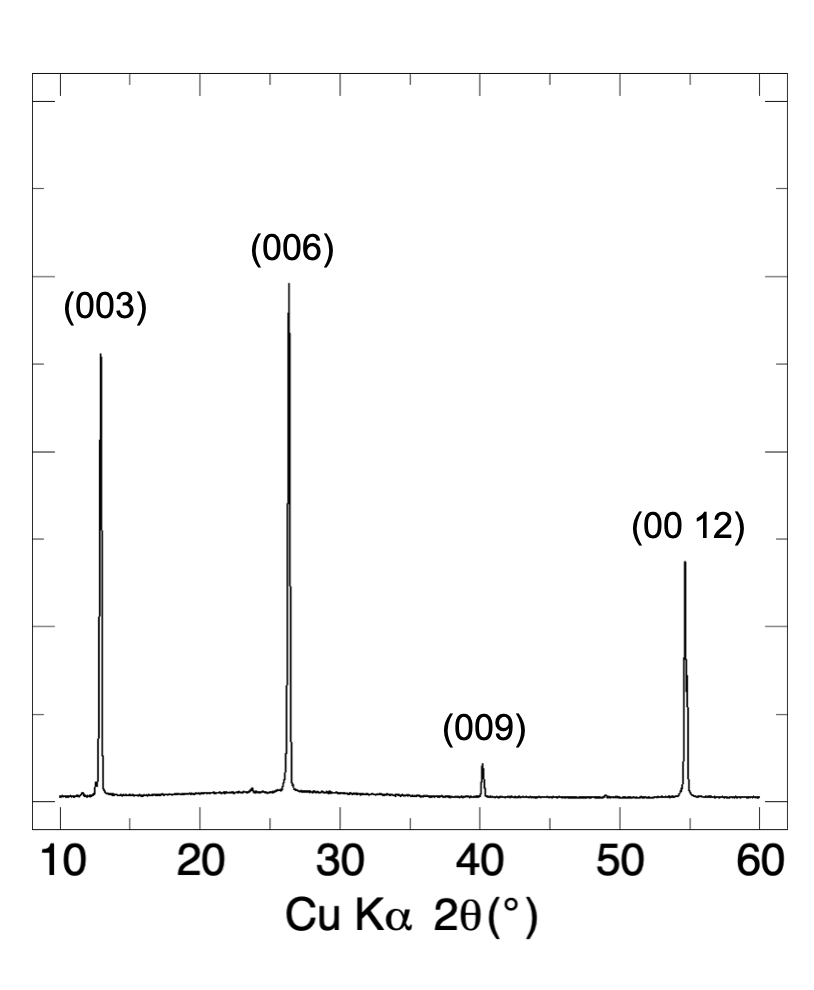}
\caption{\textbf{X-ray diffraction scan of \ch{MnPSe_3}}. 
}
\label{fig:XRD}
\end{figure*}

\section{Phonons}
\begin{figure*}[th]
\includegraphics[width=.8\textwidth]{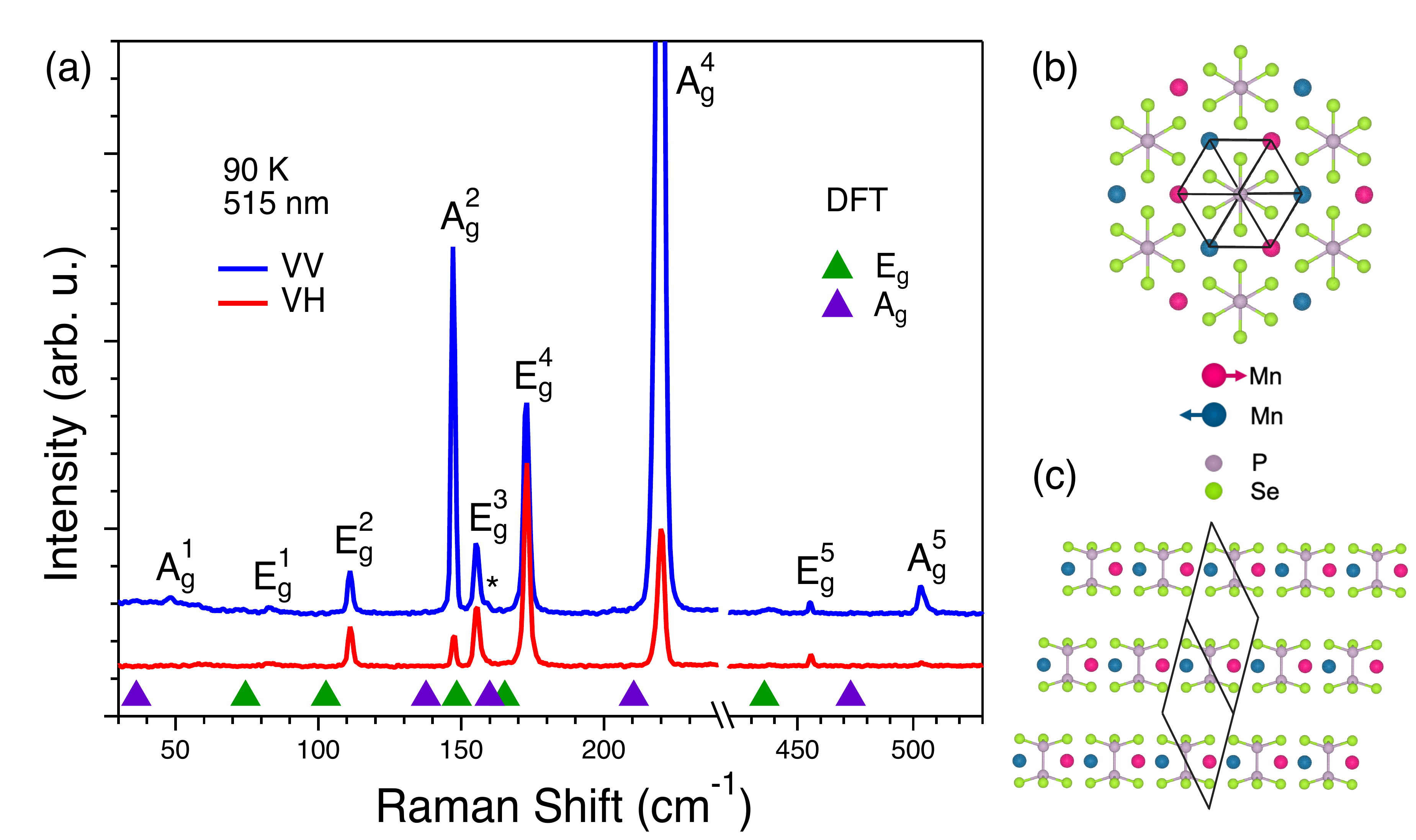}
\caption{\textbf{Raman phonons of \ch{MnPSe_3}}. (a) The Raman spectra for parallel (VV) and crossed (VH) polarizations taken with 515\,nm excitation laser at 90\,K, above \TN. The phonon modes are labeled with the irreducible representation of the point group $\overline{3}$. The * marks the weak \Ag{3} phonon. Below 74\,K, a single layer of \ch{MnPSe_3} with N\'eel type antiferromagnetic ordering are shown perpendicular to the plane (b) and along the plane (c). The $S=\frac{5}{2}$ Mn ions point mostly parallel to the van der Waals plane. The solid gray lines represent the rhombohedral primitive cell. The crystal lattice visualizations were made by VESTA\cite{VESTA}. 
}
\label{fig:Phonons}
\end{figure*}

\begin{figure*}[th]
\includegraphics[width=.65\textwidth]{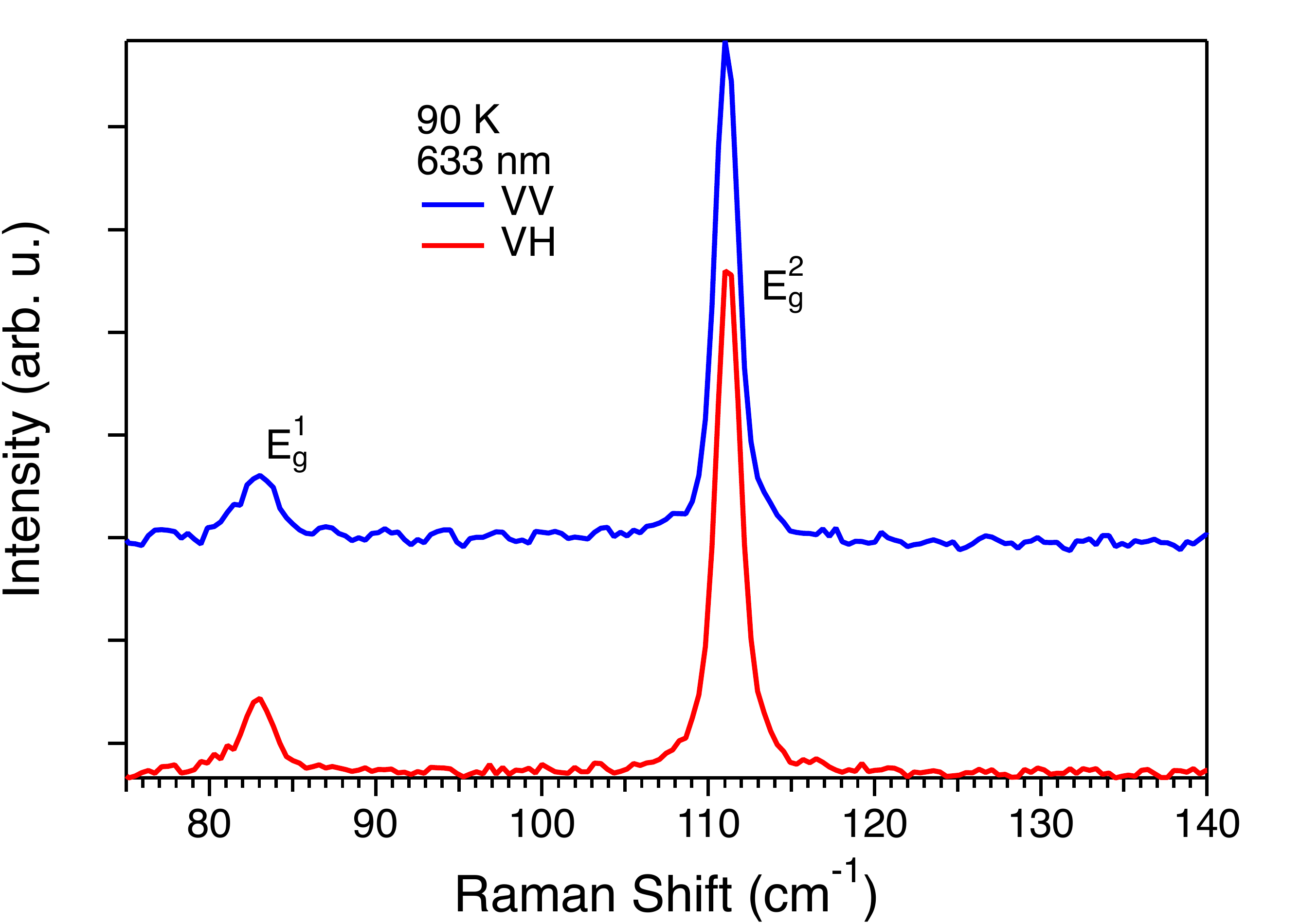}
\caption{\textbf{Select Raman phonons with 633\,nm excitation}. \Eg{1} and \Eg{2} Raman phonon at 90 K, in parallel (VV) and crossed (VH) polarization configuration.
}
\label{fig:Eg633}
\end{figure*}

The Raman spectra above \TN, at 90 K, are shown with parallel (VV) and crossed (VH) polarization configuration in SI-Fig \ref{fig:Phonons}a. In addition, the DFT results are compared to the experimental spectra to assign each observed peak with its phonon symmetry according to the $\overline{3}$ point group. We find that the DFT frequencies are consistently $\approx$ 10\,\% below the experimental peaks, which is a typical level of agreement. The full Raman phonon symmetry assignments and the peak frequencies are tabulated in Table \ref{tab:RamanModes}. We confirm the \Eg{1} phonon frequency by using an excitation wavelength of 633\,nm from a HeNe laser (SI-Fig. \ref{fig:Eg633}).

\renewcommand{\arraystretch}{1.5}
\begin{table*}[ht]
\begin{tabular}{|W{c}{45pt}|W{c}{70pt}|W{c}{70pt}|W{c}{70pt}|}
\hline
Symmetry      &  DFT (\wn)   &   Exp. (\wn)  &  Diff. (\wn) \\ \hline
\Eg{1}        &  74.3		  &   84.0		 &	9.7			\\ \hline
\Eg{2}		  &	 102.4		  &	  109.6		 &	7.2			\\ \hline
\Eg{3}		  &	 148.0		  &	  156.9		 &	8.9			\\ \hline
\Eg{4}		  &	 164.7		  &	  174.5		 &	9.8			\\ \hline
\Eg{5}		  &	 436.4		  &	  455.6		 &	19.2		\\ \hline
\Ag{1}		  &	 36.2		  &	  48.3		 &	12.1		\\ \hline
\Ag{2}		  &	 137.3		  &	  149.0		 &	11.7		\\ \hline
\Ag{3}		  &	 159.5		  &	  162.2		 &	2.7			\\ \hline
\Ag{4}		  &	 209.8		  &	  222.0		 &	12.2		\\ \hline
\Ag{5}		  &	 473.4		  &	  506.0		 &	32.6		\\ \hline
\end{tabular}
\caption{\textbf{List of $\mathbf{E_g}$ and $\mathbf{A_g}$ phonon frequencies in units of cm$^{-1}$.} The experimental results (at 90\,K) is compared with calculated values from DFT.}
\label{tab:RamanModes}
\end{table*}
\renewcommand{\arraystretch}{1}

\begin{figure*}[th]
\includegraphics[width=0.7\textwidth]{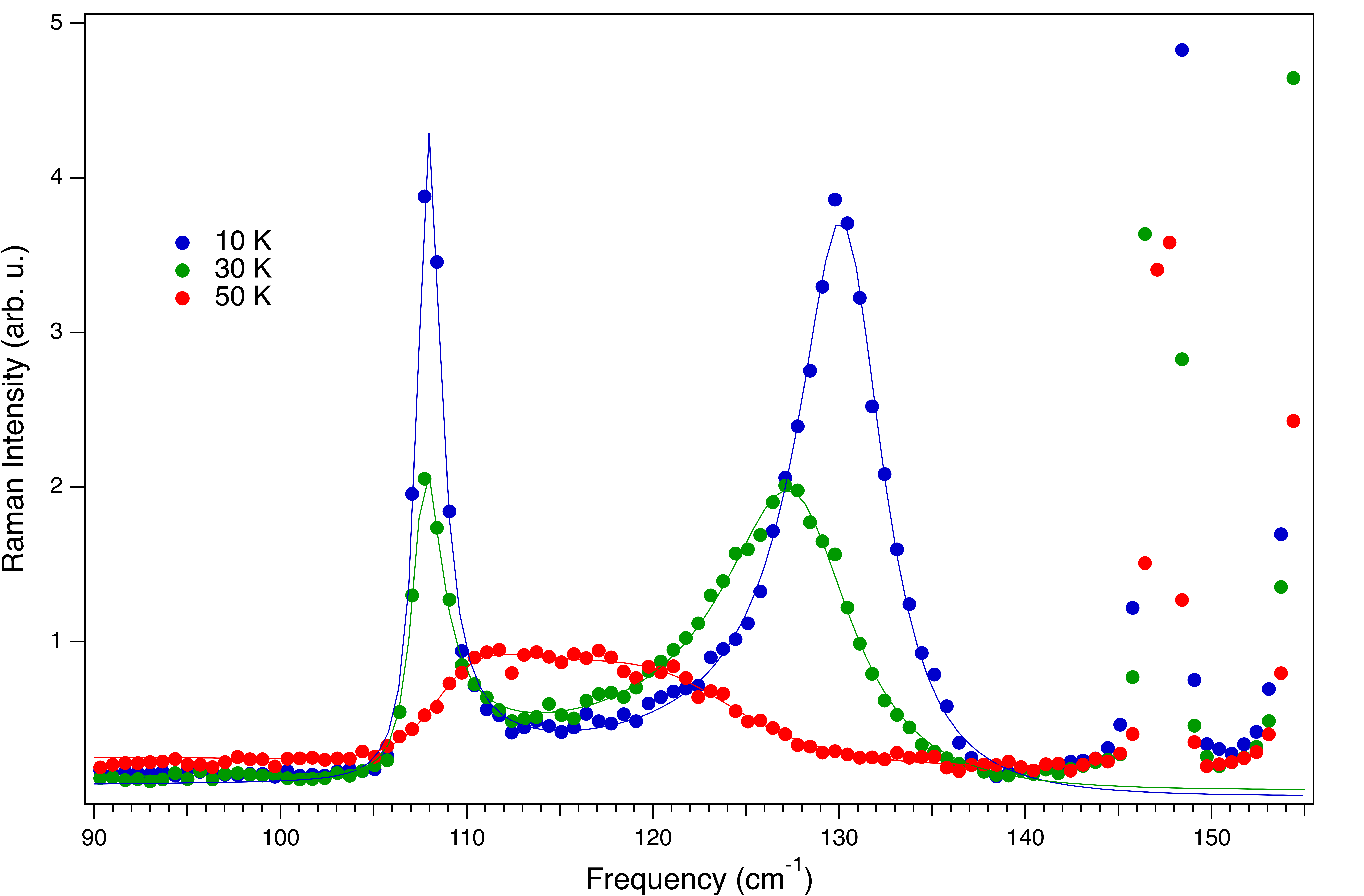}
\caption{\textbf{Examples of double Fano fit} The solid lines are the best fit of the data (solid circle) between 100\wn\ to 140\wn. 
}
\label{fig:FanoExample}
\end{figure*}

The frequency range between 100\wn\ to 140\wn\ are fitted with a function that comprises of 2 Fano lineshapes. Here, we choose to use a normalized Fano lineshape multiplied by a constant A, for area. Some fit examples are shown in SI-Fig \ref{fig:FanoExample}.

\begin{equation}\label{fitfunc}
f_{Fano} (\omega,q,\Gamma) = \frac{A}{1+q^2} \frac{(\frac{2(\omega-\omega_0)}{\Gamma}+q)^2}{(\frac{2(\omega-\omega_0)}{\Gamma})^2+1}
\end{equation}

The other phonons are fitted with the Voigt profile.

\begin{figure*}[h]
\includegraphics[width=0.5\textwidth]{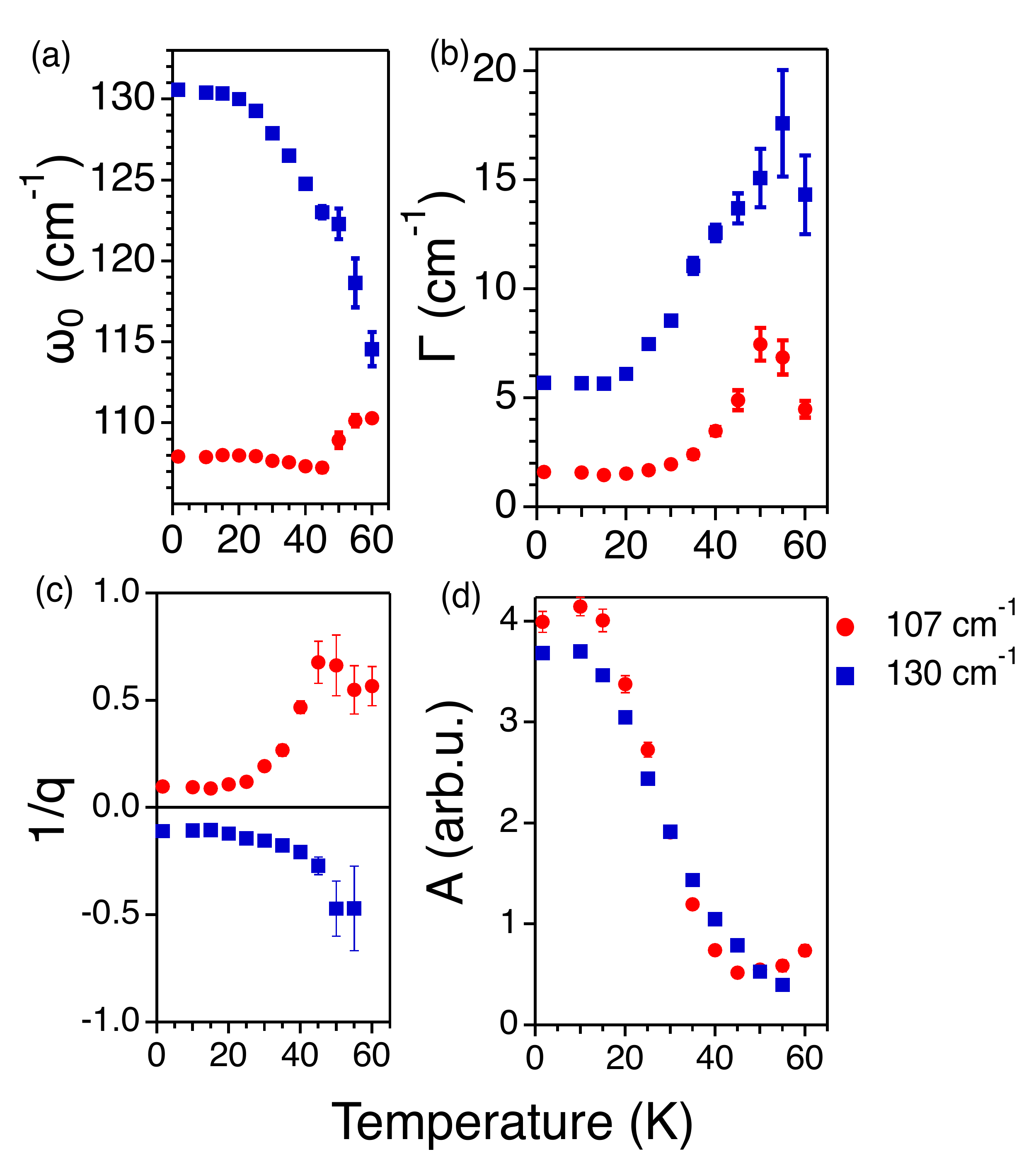}
\caption{\textbf{Results from double Fano fit}. 
}
\label{fig:FanoResults}
\end{figure*}

The result of fitting the two Fano peaks is shown in SI-Fig. \ref{fig:FanoResults}. The lifetime in the main text is approximated by taking the inverse of the spectral width of the modes ($\Gamma$). The value $q$ represents the asymmetry of the linewidth, which is a function of the interaction and energy overlap of the discrete mode and the continuum \cite{Fano1961}. As $q\rightarrow\infty$, (or $\frac{1}{q} \rightarrow 0$), the function becomes more Lorentzian. The temperature dependence of the area (SI-Fig \ref{fig:FanoResults}d) has been attributed to the spin-dependent Raman scattering process\cite{Makimura_1993}. 

\begin{figure*}[h]
\includegraphics[width=0.5\textwidth]{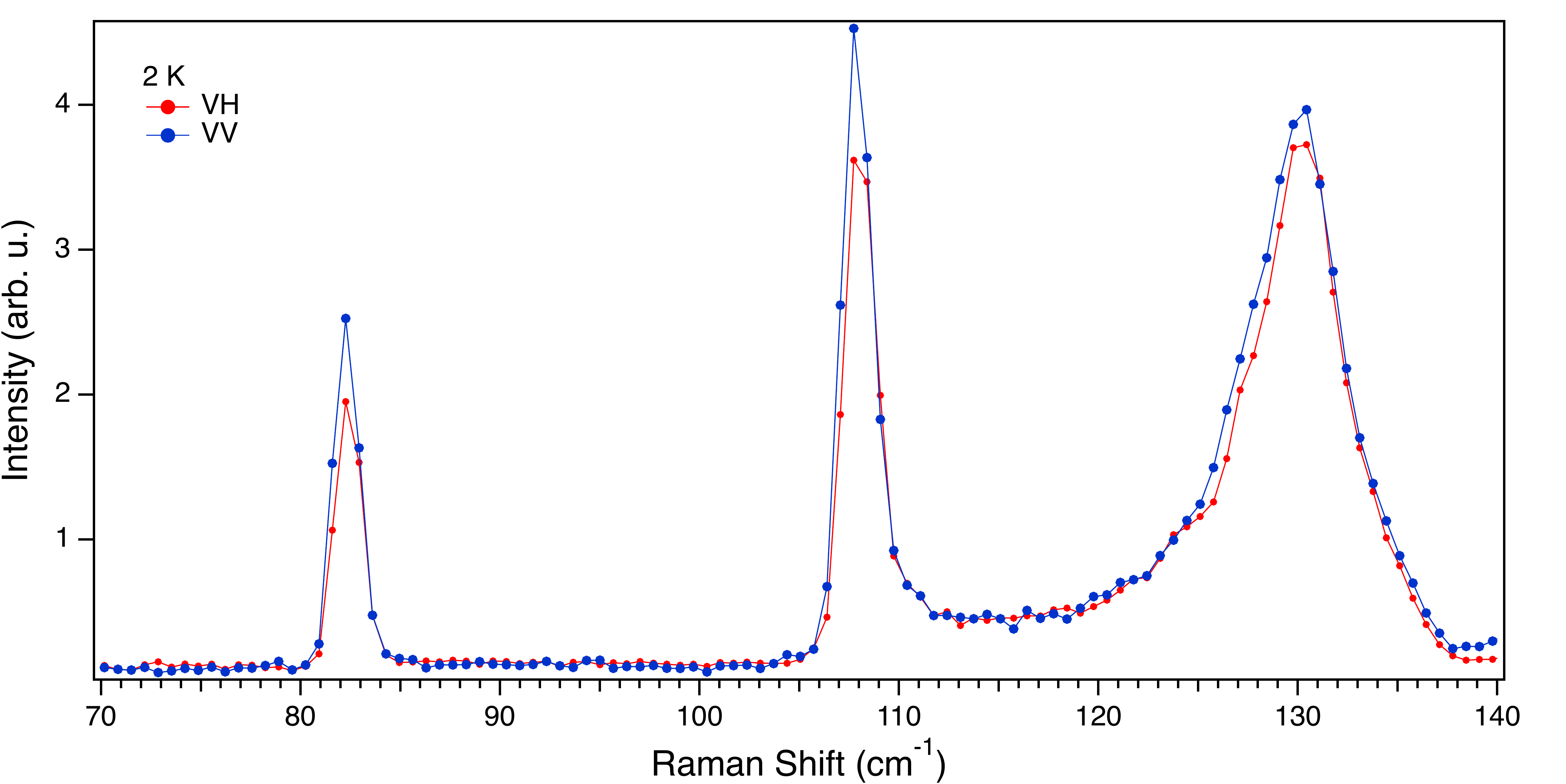}
\caption{\textbf{Comparison between parallel and crossed polarization configurations}. 
}
\label{fig:ParallelCross}
\end{figure*}

\section{Magnetic field dependent}

\begin{figure*}[th]
\includegraphics[width=0.6\textwidth]{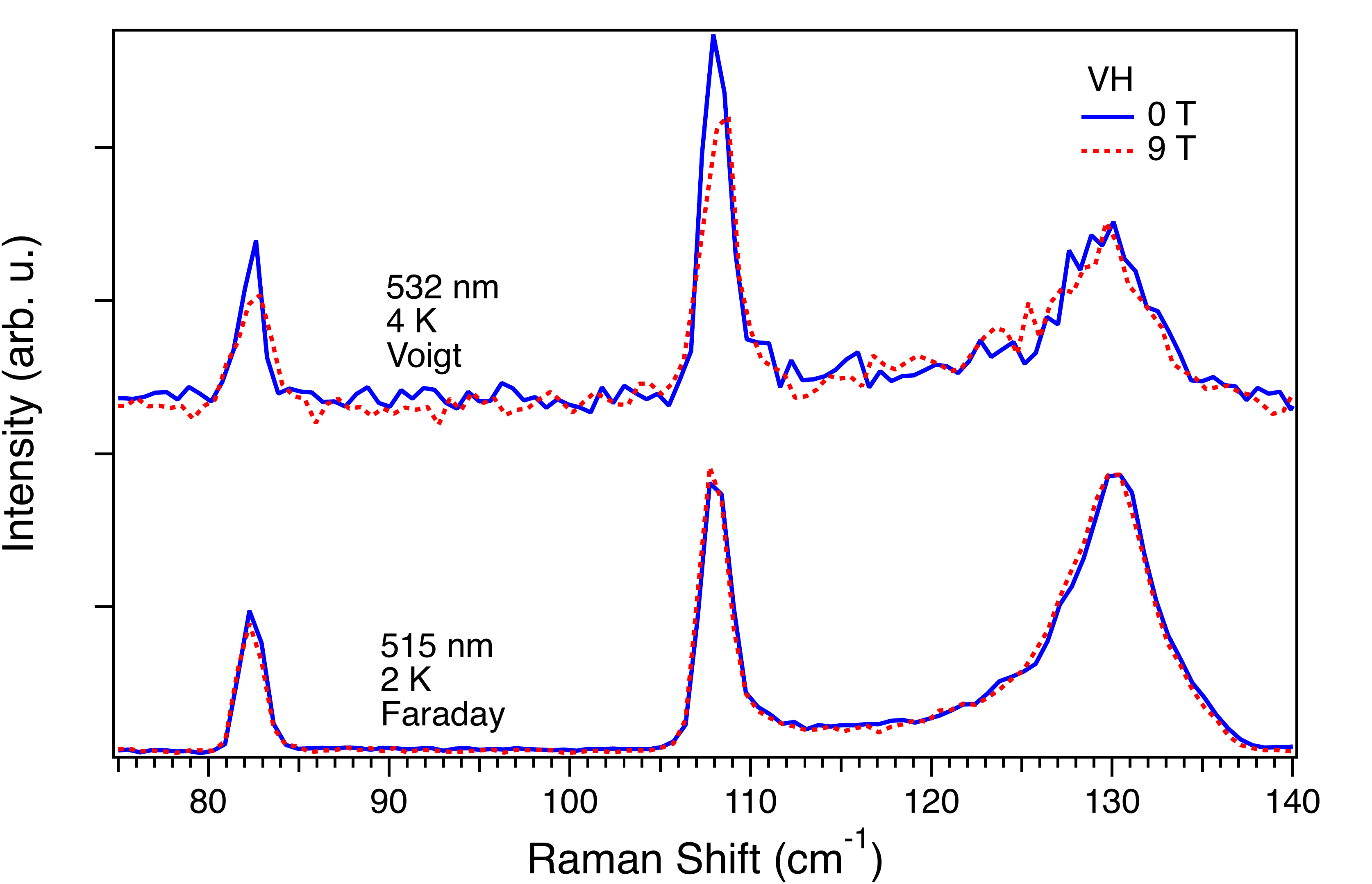}
\caption{\textbf{Comparison between magnetic field dependent}. In crossed polarization configuration, between Faraday (applied magnetic field perpendicular to the layer) and Voigt (applied magnetic field parallel to the layer), no change can be seen in the peak at 130\wn between 0\,T and 9\,T, for multiple excitation laser wavelengths, 515\,nm and 532\,nm.
}
\label{fig:FieldDepend}
\end{figure*}
An external static magnetic field can be applied perpendicular to the van der Waals plane (Faraday geometry) or parallel to it (Voigt). We see no change in the scattering intensity around 130\wn\ between 0\,T and 9\,T.

\section{DFT and LSWT calculations and Neutron Comparison}

\begin{figure*}[th]
\includegraphics[width=0.6\textwidth]{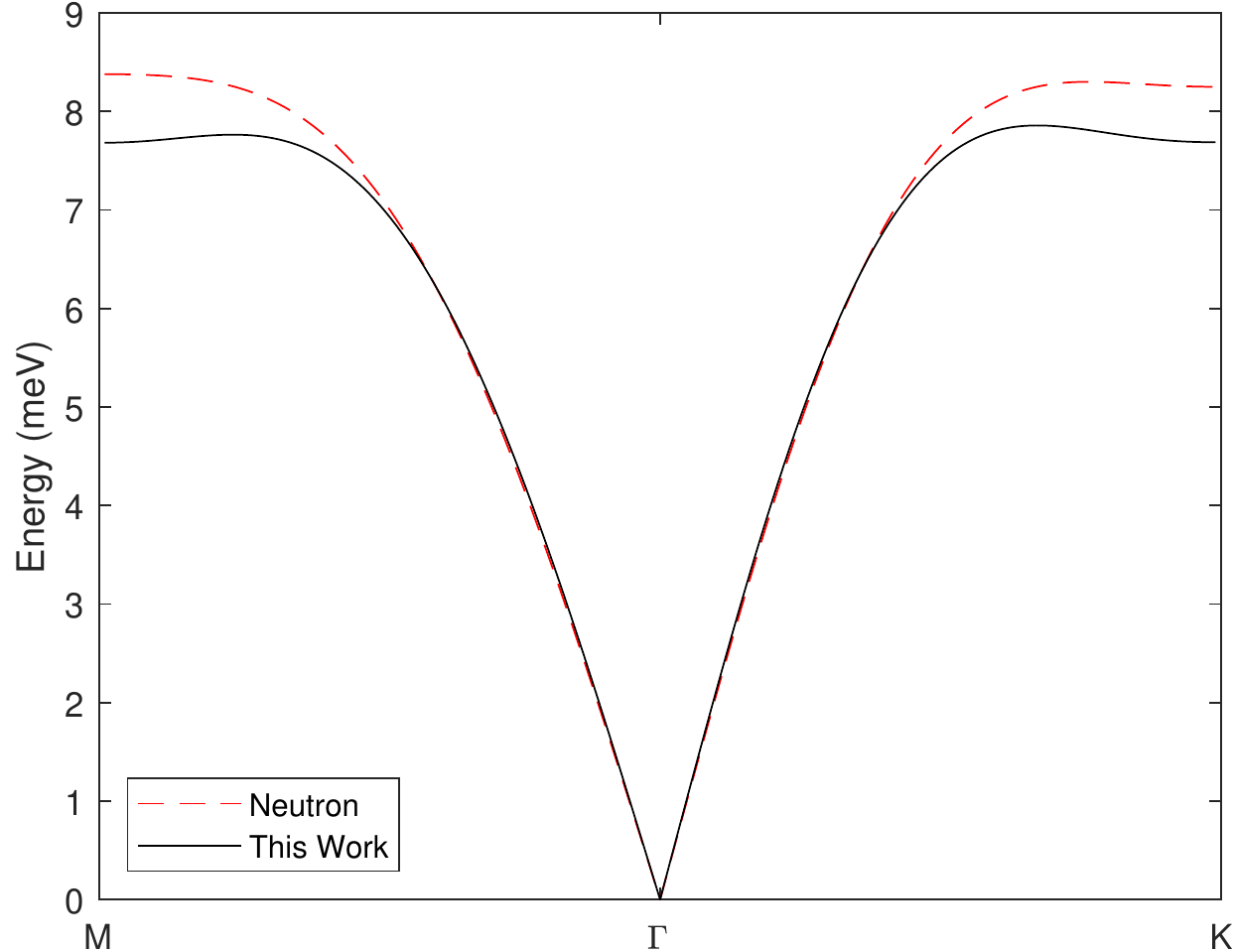}
\caption{Comparison of the magnon spectrum in this work, based on DFT+U with U=4 eV, with the Neutron scattering fit of \cite{calder2020magnetic}.
}
\label{fig:magcomp}
\end{figure*}

\begin{figure*}[th]
\includegraphics[width=0.6\textwidth]{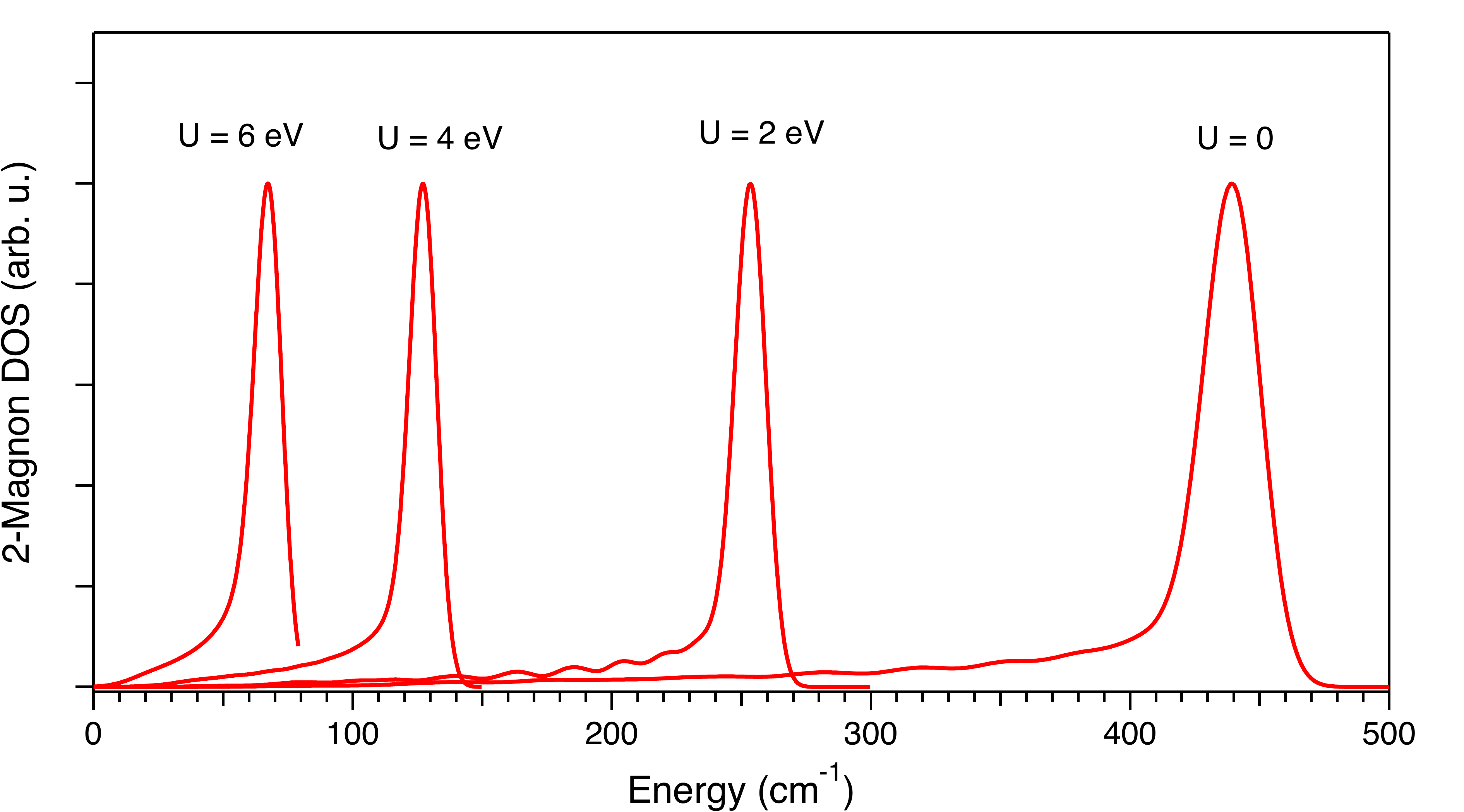}
\caption{\textbf{Comparison of the 2-magnon density of states for different values of Hubbard U}. U = 4\,eV was chosen in the main text to match the experimental peak around 130\wn.
}
\label{fig:MagDOS2}
\end{figure*}

Very recently, there has been elastic and inelastic neutron scattering performed on MnPSe$_3$\cite{calder2020magnetic}. In that work, they fit a Heisenberg model to the magnon specturm by keeping the ratios of the in-plane $J$ Heisenberg couplings fixed to theoretical work\cite{Sivadas_2015,Yang_2020} and fiting an overall value of the in-plane coupling constants. They also fit a single effective out-of-plane coupling, and do not include any anisotropic terms. In SI-Fig.~\ref{fig:magcomp}, we show a comparison between that work and our DFT-based in-plane magnon spectrum, without including the anisotropy and with U=4 eV. We find excellent agreement overall, with small disagreement at the band edges. However, as shown in SI-Fig.~\ref{fig:MagDOS2}, the band edge of the magnon spectrum is extremely sensitive to the value of U. Therefore, within the accuracy of the underlying DFT+U theory, we get excellent agreement with the experimental spectrum.

\section{Phonon - Two-magnon continuum interaction model}

We model the interaction of a discrete phonon resonance with a two-magnon continuum. For small coupling, we use the formalism of Fano\cite{Fano1961}. Specifically, Eqs. 2 and 13 of that work, 
\begin{eqnarray}
F(E) &=& P \int dE' \frac{|V_{E'}|^2}{E - E'} \label{eq:f}\\
|a(E)|^2 &=& \frac{|V_E|^2}{(E - E_{\phi} - F(E))^2 + \pi^2 |V_E|^4}
\end{eqnarray}
where $E_{\phi}$ is the energy of the discrete resonance, $V_E$ is the coupling between the resonance and the continuum at energy $E$, $P$ stands for the principle value, $F(E)$ is the shift in the resonance location, and $|a(E)|^2$ is proportional to the Raman signal. 

We choose a simple functional form for $|V_E|^2$ by using two line segments that form a triangular peak, which approximate the shape of the two-magnon continuum. 
\begin{equation}
|V_E|^2 = \begin{cases} 
0 & E < E_L \\
t^2 (E - E_L)/(E_P-E_L) & E_L < E < E_P \\
t^2 (E_U - E)/(E_U-E_P) & E_P < E < E_U \\
0 & E > E_U \end{cases}
\end{equation}
where $E_L$ is the lower energy of the two-magnon continuum, $E_U$ is the upper energy, $E_P$ is the two-magnon peak position, and $t$ is a coupling constant. Then, we can perform the integral in Eq.~\ref{eq:f} and get frequency shift of the peak:
\begin{eqnarray}
F(E) &=& F_1 + F_2 \\
F_1 &=& t^2 Real[\; \frac{E - E_L}{E_P - E_L} log(\frac{E_P - E}{E_L - E}) \;] \\
F_2 &=& t^2 Real[\; \frac{E_P - E}{E_U - E_P} log(\frac{E_U - E}{E_P - E}) \;]
\end{eqnarray}
This shift extends beyond the region that the discrete resonance directly overlaps with the two-magnon continuum. The extra line-width of the phonon resonance due to the interaction (beyond the intrinsic phonon line-width) is proportion to $\pi |V_E|^2$. Unlike the resonance shift, the broadening is only non-zero when the phonon overlaps with the continuum.

This model can also be solved numerically for arbitrary coupling by considering a Hamiltonian with a single phonon resonance that couples to a large number of states that approximate a continuum.
\begin{eqnarray}
H = \begin{pmatrix} E_\phi & t_1 & t_2 &  t_3  & ...\\
t_1  & E_1  & 0  & 0  & ...\\
t_2  & 0 &  E_2  &  0 &  ...\\
t_3  & 0  & 0  & E_3  &  ... \\
... \end{pmatrix}
\end{eqnarray}
Here, the $E_i$ are energies approximating the continuum, and $t_i$ is the coupling between the continuum at $E_i$ and the discrete phonon ($E_\phi$). The system can easily be solved numerically for a few hundred continuum states, getting eigenvalues $\epsilon_i$ and eigenvectors $v_i$. The phonon resonance will be proportional to
\begin{eqnarray}
|a(E)|^2 \propto \sum_i \frac{|v^\phi_i|^2}{(\epsilon_i - E)^2 + \Gamma_{ph}^2}
\end{eqnarray}
where $v^\phi_i$ is the phonon component of the $i$-th eigenvector, and $\Gamma_{ph}^2$ is the intrinsic phonon linewidth. The results agree with the formalism above for weak coupling.

\bibliography{MnPSe3References}